\newcommand{\rev}[1]{\textcolor{black}{#1}}

\documentclass[sigconf]{acmart}

\AtBeginDocument{%
  \providecommand\BibTeX{{%
    \normalfont B\kern-0.5em{\scshape i\kern-0.25em b}\kern-0.8em\TeX}}}

\setcopyright{acmlicensed}
\copyrightyear{2025}
\acmYear{2025}
\acmDOI{XXXXXXX.XXXXXXX}

\acmConference[]{}{}{}
\setcopyright{none}
\renewcommand\footnotetextcopyrightpermission[1]{}

\begin{document}

\title{Choosing the Right Git Workflow: A Comparative Analysis of Trunk-based vs. Branch-based Approaches}


\author{Pedro Lopes}
\affiliation{%
  \institution{Centro de Informática}
  \institution{Universidade Federal de Pernambuco}
  \country{Recife, Pernambuco, Brazil}}
\email{phls2@cin.ufpe.br}

\author{Paola Accioly}
\affiliation{%
  \institution{Centro de Informática}
  \institution{Universidade Federal de Pernambuco}
  \country{Recife, Pernambuco, Brazil}}
\email{prga@cin.ufpe.br}

\author{Paulo Borba}
\affiliation{%
  \institution{Centro de Informática}
  \institution{Universidade Federal de Pernambuco}
  \country{Recife, Pernambuco, Brazil}}
\email{phmb@cin.ufpe.br}

\author{Vitor Menezes}
\affiliation{%
  \institution{Centro de Informática}
  \institution{Universidade Federal de Pernambuco}
  \country{Recife, Pernambuco, Brazil}}
\email{vitorcardimmenezes@gmail.com}


\begin{abstract}
Git has become one of the most widely used version control systems today. Among its distinguishing features, its ability to easily and quickly create branches stands out, allowing teams to customize their workflows. In this context, various formats of collaborative development workflows using Git have emerged and gained popularity among software engineers. We can categorize such workflows into two main types: branch-based workflows and trunk-based workflows. Branch-based workflows typically define a set of remote branches with well-defined objectives, such as feature branches, a branch for feature integration, and a main branch. The goal is to migrate changes from the most isolated branch to the main one shared by all as the code matures. In this category, GitFlow stands out as the most popular example. In contrast, trunk-based workflows have a single remote branch where developers integrate their changes directly. In this range of options, choosing a workflow that maximizes team productivity while promoting software quality becomes a non-trivial task. Despite discussions on forums, social networks, and blogs, few scientific articles have explored this topic. In this work, we provide evidence on how \rev{Brazilian} developers work with Git workflows and what factors favor or hinder the use of each model. To this end, we conducted semi-structured interviews and a survey with software developers. Our results indicate that trunk-based development favors fast-paced projects with experienced and smaller teams, while branch-based development suits less experienced and larger teams better, despite posing management challenges.
\end{abstract}





\maketitle
\pagestyle{empty}

\section{Introduction}\label{sec:intro}

Among existing version control systems (VCS), Git stands out as an increasingly popular tool for its open-source nature and for providing both quick and easy creation of development branches, allowing the implementation of new modules and features in isolation without affecting the primary code repository. This functionality promotes modular software development and facilitates task division. Thus, development teams can customize how they want to collaborate.

In this context, various workflows with Git became popular among software engineers/developers. In this work, we categorize such workflows into the following: \textit{branch}-based and \textit{trunk}-based. While branch-based workflows organize the software development cycle into well-defined stages using separate branches, trunk-based workflows integrate all code directly into the main branch. 

GitFlow is the most popular example of a branch-based workflow. It consists of guidelines for creating five different branches according to the maturity level of the code and the type of task that is implemented \cite{CORTESRIOS}. The main idea is that developers work on new features and bug fixes in isolated branches. After implementation, they migrate changes from the most isolated branch until they reach the main branch, which the team shares, as the code is verified and validated. Although GitFlow is the most popular example, each company can define its workflow with other branch structures, such as mirroring system architecture, team structure, or both \cite{SHIHAB}. 

On the other hand, in trunk-based workflows, the team commits their changes to the only main branch, the trunk, multiple times a day, without extensive branch creation, ensuring that the codebase is always available on demand \cite{TRUNK}. 

In this range of options, choosing a workflow that maximizes team productivity while promoting software quality becomes a non-trivial task. Despite the discussion in online forums, social networks, and blogs, few scientific studies explore this issue in depth: some focused on branching structures and the effects of excessive branching practices \cite{SHIHAB, BIRD, COSTA, PHILLIPS}, others extracted Git workflows characteristics from public repositories \cite{CORTESRIOS}, and studies even compared both types of workflow \cite{DEBOER, NEELYSTOLT}, \rev{but they focused on a specific development context: companies switching from one workflow to another}.

\emergencystretch 3em
In this work, we conducted both online surveys and semi-structured interviews to understand how developers work collaboratively using different Git workflows in their daily activities and assess their perception of which factors encourage or hinder the usage of one workflow over the other. \rev{In total, we conducted 22 interviews and had 54 responses to our survey.}

Providing this kind of evidence is necessary because it offers an overall perspective of what developers are using, best practices, potential pitfalls, and the identification of specific contexts where one workflow might perform better than the other. This knowledge can aid companies, particularly those in their early stages, in adopting reported best practices or deciding which workflow to use based on their development context, contributing to future research, such as developing new Git workflow guidelines or automation tools.

\rev{Based on our sample,} our results indicate that branch-based workflows are more widely known and used in practice than trunk-based ones \rev{among Brazilian developers.} Also, while trunk-based workflows favor fast-paced projects with experienced developers and smaller teams, branch-based development suits less experienced and larger teams better, despite posing management challenges due to all the steps developers need to follow to deliver their code.

Moreover, due to ethical research concerns involving human subjects, we submitted this study to Plataforma Brasil, a unified national database of research records involving human subjects, which was approved \rev{(approval no. 6.740.037).}
\section{Study setup}\label{sec:methodology}
The goal of this work is to learn from software developers which workflows are used the most in practice, considering branch-based and trunk-based categories, how they perceive the advantages and disadvantages of each type, and whether they understand which factors (technical, organizational, or social) favor or hinder the adoption of each workflow. Based on their reality, this research will generate recommendations for individuals unsure of which workflow to adopt.

To achieve this goal, we conducted an online survey and semi-structured interviews to answer the following research questions. 

\begin{itemize}
    \item RQ1. How do developers work with Git Workflows?
    \item RQ2. What factors favor or hinder using a branch-based or a trunk-based workflow?
\end{itemize}

We adopted a mixed methods approach that combined both survey and interviews to better understand developers' perspectives. The survey provided an initial quantitative overview that helped identify general trends and patterns, while the interviews allowed deeper qualitative insights, capturing contextual details and nuances that the survey alone could not provide.

To address \textbf{RQ1}, both through the survey and the semi-structured interviews, we used the following question:

\emph{Regarding the project you are currently working on (if you work on more than one currently, please choose one), describe the \textbf{step-by-step} process of a code change from the moment you implement it until it is pushed to the remote repository and subsequently deployed into production.}

Our objective with this question is to compute the frequency of usage for each category. Besides that, we would like to delve into the step-by-step process of each workflow to identify potential differences among the various possible formats within each category, particularly in branch-based workflows, which can be highly customizable.

Regarding \textbf{RQ2}, we did not want to bias the responses of the participants with our research hypotheses regarding factors that favor or hinder a specific workflow. Therefore, we asked the questions in the most generic way possible, inquiring about which factors favor or hinder branch-based and trunk-based workflows without mentioning any specific factors.

In addition, we have crafted some questions that would help better characterize our sample, such as how many years of experience the respondent has in software development activities, the size of their team, and their current role. These data are also helpful in establishing potential correlations, such as whether there is a correlation between the chosen workflow and the participant's experience, among others.

Lastly, we also ask whether respondents want to change their current workflow and why. This question would provide another opportunity to understand the participants' perceptions of the factors they consider relevant for decision-making.

After preparing the initial version of the interview script and the questionnaire for the survey, we piloted both studies with the first participant each to test whether the instruments were suitable.

In the case of the interviews, the participant suggested two new questions, which we added to the remaining interviews. The first question regarded whether the participant classified the company they worked for as a startup, and the second asked if the participant encountered recurring issues with merge conflicts.

Despite these questions mentioning specific factors, we believe that they do not bias responses to the generic questions about factors cited earlier, as we ask them before we mention them. In addition, such questions help provide better context about the interviewee's context.

It is important to mention that the survey and interviews occurred independently from this initial jointly conceived version onward. For this reason, some questions underwent slight alterations after their pilot round, resulting in minor differences. Therefore, in Section \ref{cap:results_and_discussion}, there are some discussions in which we will describe graphs with different metrics for both studies.

For example, in the survey, we asked the respondents how many years of experience they had with software development activities. In contrast, in the interviews, we asked about the interviewee's current position, ranging from junior, mid-level, to senior.

\subsection{Interviews and survey application}

For the interviews, we initially selected 37 participants from the authors' contact network, prioritizing engineers/developers with a higher seniority level and who still work directly with code rather than managing teams. The rationale behind this is that code management is performed and affects more developers than managers, as they are the ones dealing with it daily. After this selection process, we interviewed 22 individuals. All instruments used for both survey application and interviewing, as well as interview transcriptions in Portuguese, are available online in our artifact package \cite{GIT_WORKFLOWS_REPO}.

Regarding the survey, we shared our online questionnaire on social networks like LinkedIn, Instagram, WhatsApp, and Twitter. In addition to that, we promoted our questionnaire through email lists, with eventual reminders to increase the chances of obtaining more responses. In total, we obtained 54 responses, but we excluded four answers, as participants claimed that they did not use Git or a different VCS since they occupied managerial roles.

\subsection{Data analysis}\label{sub:data_analysis}
To answer \textbf{RQ1}, we needed to establish if the participant described a branch or trunk-based workflow. During the interview, this process was straightforward since the interviewer could ask for more details. However, some of the answers in our survey were difficult to analyze, as they were more succinct. To make this analysis more systematic and increase our confidence in the collected data, we obeyed the following rules to classify a workflow as branch or trunk-based:

\begin{enumerate}
    \item If the participant explicitly cites the type of workflow in his description,  we would classify it in its respective category;
    \item We categorize the answer into the branch-based group if a workflow has multiple remote branches where developers collaborate, like develop and feature branches, and as trunk-based otherwise;
    \item  If the description were confusing, we would discuss it in pairs to check if we could reach a consensus;
    \item If we could not reach a consensus, we would classify the workflow in the "unidentified" category.
\end{enumerate}

To answer \textbf{RQ2} using the interviews' transcriptions as input, we followed the Saldana~\cite{saldana} and Vasilescu~\cite{youtube} approach to encode and analyze the data, opting for an open coding technique. We used the first interview as a pilot to create initial codes, and the subsequent interviews as supplementary to complete the coding process. We employed a flexible approach and adjusted the codes as needed, as we describe in the following section.

To analyze the survey responses, we adopted the card sorting approach inspired by Zimmermann~\cite{cardsorting} to identify common themes in the open questions. First, we extracted main topics from each response regarding factors that favor or hinder workflows and created cards for them. When closely related topics appeared, like "ease of use" and "lower complexity", which convey the same idea, we grouped them, repeating the process until we addressed every answer. Lastly, we named the groups with the factor that best represented their cards. 
\section{Results and Discussion}\label{sec:results}
\label{cap:results_and_discussion}
In this Section, we describe the demographic characteristics of our sample and summarize the responses provided for \textbf{RQ1} and \textbf{RQ2}.

\subsection{Demographics}
\label{sec:demographics}
Overall, we interviewed 22 people and received 54 responses to our survey, all from Brazilian developers. However, we cannot specify whether the participants worked for companies only in Brazil or abroad. Moreover, we excluded four responses from our survey, as participants claimed that they did not use Git or a different VCS since they occupied managerial roles.

\subsubsection{Years of Experience}
\label{sub:yoe}
\hfill\

\begin{figure}[hbt!]
  \centering
  \caption{\label{fig:experience} Survey Developers' Years of Experience}
  \includegraphics[width=\linewidth]{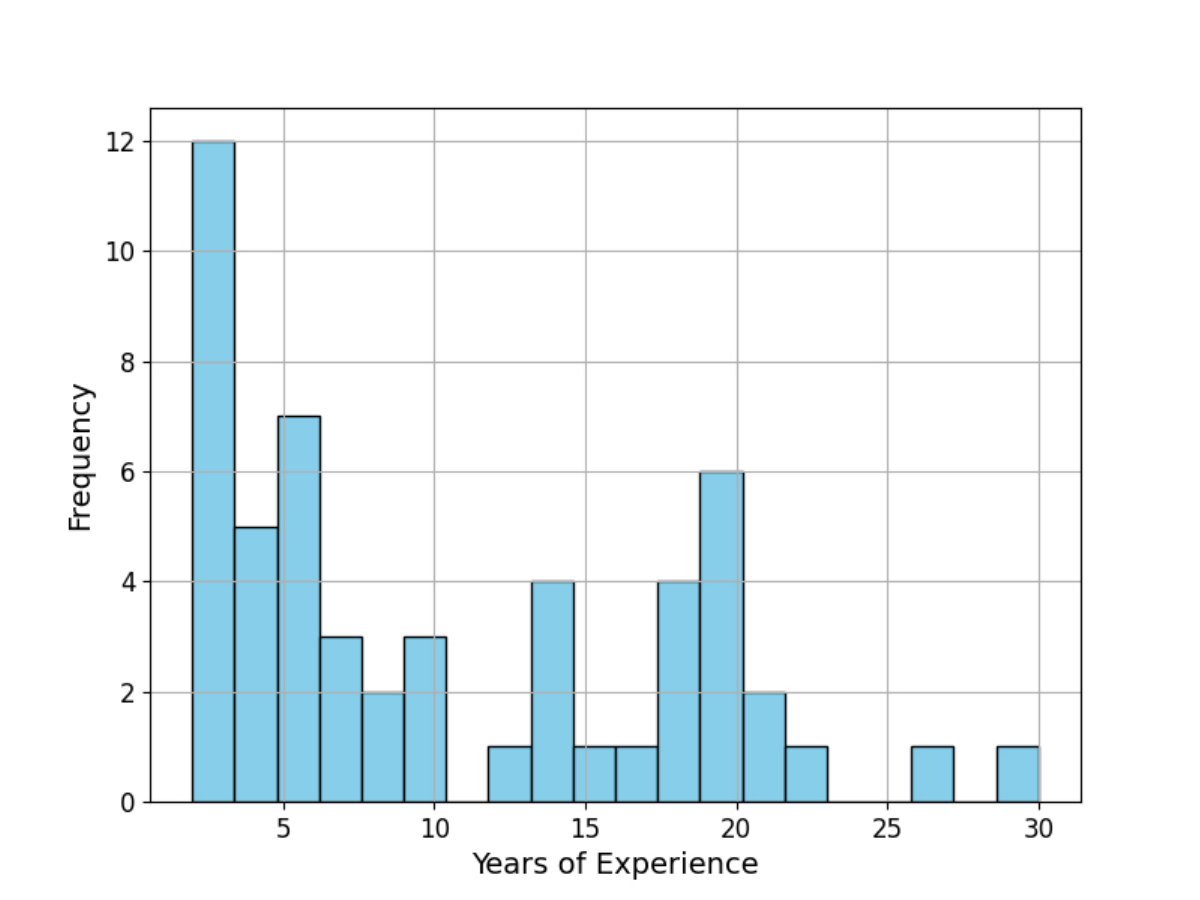}
  \Description{A histogram graph showing the years of experience from the respondent group of our survey}
\end{figure}

Among the questions used to characterize our sample, in the survey, we asked about the number of years of experience of the participants. In contrast, we asked about the interviewee's current position in the interviews. The histogram in Figure \ref{fig:experience} presents data from years of experience, and a discussion on the latter can be seen in Subsection \ref{sub:jobtitle}.

In the survey, the Mean ($M$) years of experience is 10.56, the Median ($Mdn$) is 7.5 and the standard deviation ($SD$) is 7.61.

In Figure \ref{fig:experience}, we can divide our sample into two groups \rev{using $Mdn$ as the pivot, as this value indicates the exact point at which 50\% of the respondents fall below and 50\% above.} The first group comprises developers with fewer years of experience (ranging between 2 and 7.5), while the other includes more experienced developers (ranging from 7.5 to almost 30 years).

The first group shows more concentrated values, whereas the other is sparser, with our sample having both young and experienced developers. Additionally, a $SD$ of 7.61 suggests a moderate variability in experience levels, which contributes to pushing this value upward, as we can see from the sparse data beyond 10.56.

\begin{figure}[hbt!]
  \centering
  \caption{\label{fig:seniority} Interviewees' Seniority Level}
  \includegraphics[width=\linewidth]{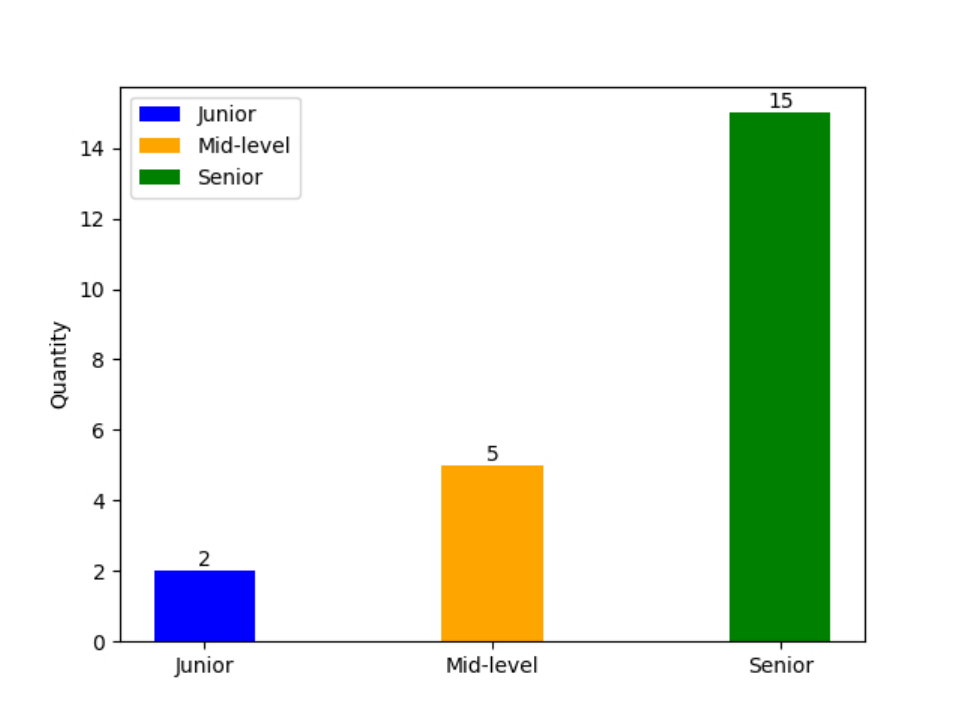}
  \Description{a bar chart describing the seniority levels of the interviewees}
\end{figure}

Figure \ref{fig:seniority}, on the other hand, shows the distribution of the current position of the respondents, ranging from the junior to senior level. As explained in Section \ref{sec:methodology}, we prioritized participants with a higher level of seniority. This is why the bar chart indicates that 75\% of our participants consider themselves senior engineers/developers. However, we also interviewed junior and mid-level developers.  

\subsubsection{Team Size}
\label{sub:teamsize}
\hfill\

\begin{figure}[hbt!]
  \centering
  \caption{\label{fig:team_size} Survey Results: Developers' Team Size Distribution}
  \includegraphics[width=\linewidth]{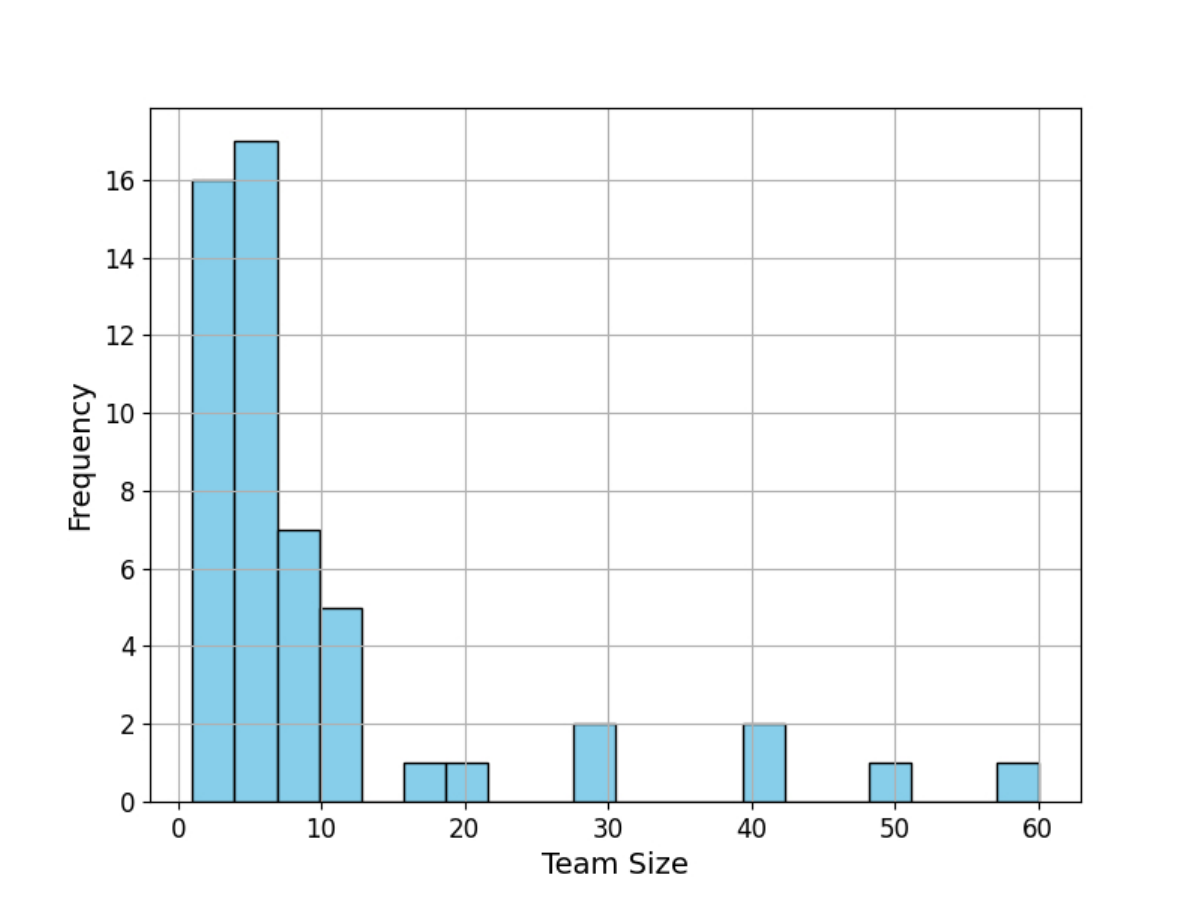}
  \Description{A histogram showing Survey Results: Developers' Team Size Distribution}
\end{figure}

\begin{figure}[hbt!]
  \centering
  \caption{\label{fig:team_sizeVitor} Interviews Results: Developers' Team Size Distribution}
  \includegraphics[width=\linewidth]{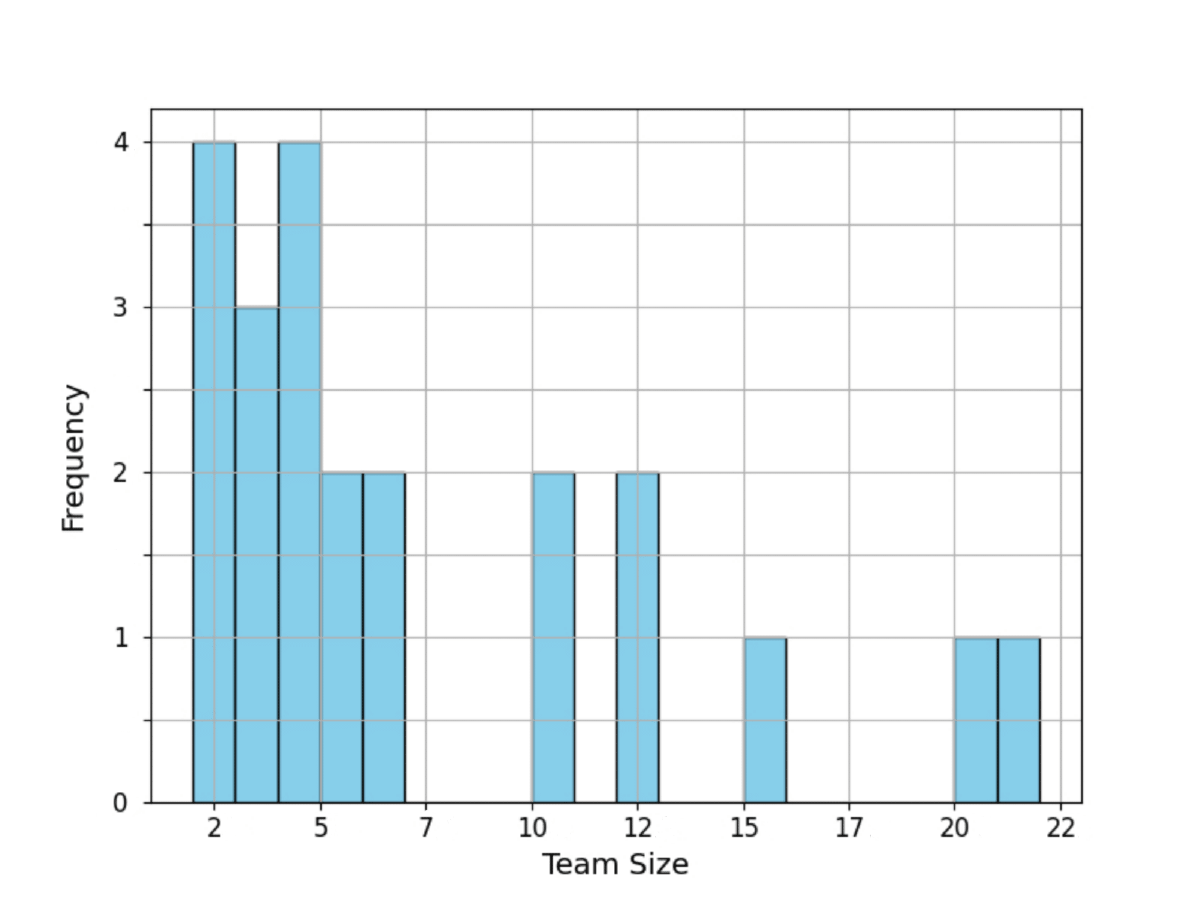}
  \Description{A histogram showing Interviews Results: Developers' Team Size Distribution}
\end{figure}

We asked both interviewees and survey respondents to report the size of the development team with which they were currently working. Figures \ref{fig:team_size} and \ref{fig:team_sizeVitor} show the histograms of the team sizes in the survey and interviews, respectively.

For the survey, the metrics calculated for the team size were as follows: our $Mdn$ is a team size of 5 developers, our $M$ is 12, and our $SD$ is 21.35. The standard deviation is high because we had an outlier with one participant indicating that their team had 137 people, which is a much higher number compared to the others. Upon reviewing this individual's responses, we found that their current position is Principal Member of Technical Staff. Therefore, they must have included all the employees below them in the company hierarchy rather than a development team.

In Figure \ref{fig:team_size}, a large majority of the answers had a team size number distributed around our $M$, with a maximum of 20 developers, with a few outliers at 30, 40, 49, 60, and 137, inflating our $M$ and $SD$ by considerable margin. Note that the biggest outlier is not represented in Figure \ref{fig:team_size}, as plotting would worsen the readability of the graph. Additionally, the most recurring value in this set was 5, appearing a total of 17 times.

Furthermore, Figure \ref{fig:team_sizeVitor} shows the histogram of the distribution of team sizes among our interviewees. For this set, $M$ is 7.09, $Mdn$ is 4.5, and $SD$ is 5.72. The most recurring values for this set are 2 and 4, which appear 4 times each.

Therefore, considering both parameters, we can conclude that participants worked in teams of different sizes; however, most of the participants in our sample (80\%) worked in teams of up to 10 people.

\subsubsection{Job Title}
\label{sub:jobtitle}
\hfill\

\begin{figure}[h!]
  \centering
  \caption{\label{fig:wordcloud_jobtitle} Survey Respondents' Job Titles Word Cloud}
  \includegraphics[width=\linewidth]{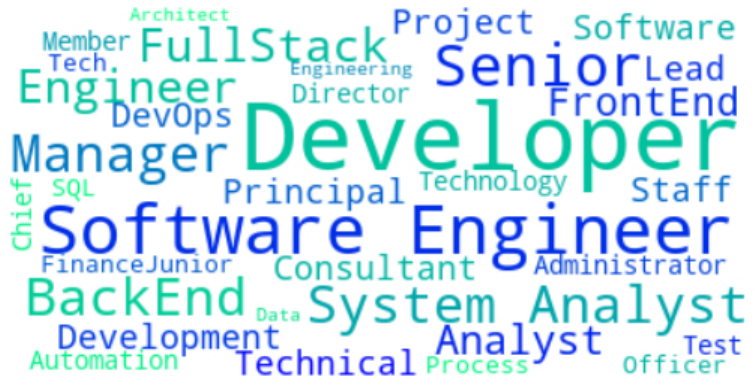}
  \Description{A word cloud graph showing the more common job titles of our sample}
\end{figure}

Along with the data provided in \ref{sub:yoe}, understanding developers' job titles can offer valuable findings into the diverse skill sets present, presenting the context for the roles, responsibilities, and areas of specialization, allowing data interpretation. In the interviews, all participants were developers/software engineers. However, to gain further insight into the distribution of job titles among survey respondents, Figure \ref{fig:wordcloud_jobtitle} shows a word cloud visualization that highlights the most prevalent titles among survey participants.

The most common job titles were Developer and Software Engineer, which appear 18 and 10 times, respectively, with Backend, Full-Stack, System Analyst, Manager, and Senior being more frequently mentioned after the first two. The Front-End job title appeared less prominent than the others. In addition, job titles involving topics such as SQL, Testing, DevOps, Automation, and Data were cited once.

Developer and Senior were big words in our word cloud because there are a lot of job titles that use this term in common with others, like Frontend Developer, Backend Developer, Senior Developer, and Senior Analyst. But in Figure \ref{fig:wordcloud_jobtitle}, we can see that our sample is diverse, mostly composed of developers from various scenarios.

\subsection{RQ1. How do developers work with Git Workflows?}
\label{sec:rq1}

To address this research question in the survey, we first compare which workflows developers currently use in their projects by analyzing their frequency based on their detailed description of a code change migrating from their local repository until it reaches the main branch. To do so, we classified the workflow into branch or trunk-based following the method described in Section \ref{sub:data_analysis}.

\begin{figure}[h]
  \centering
  \caption{\label{fig:branch_vs_trunk}Survey Results: Branch-based vs. Trunk-based}
  \includegraphics[width=\linewidth]{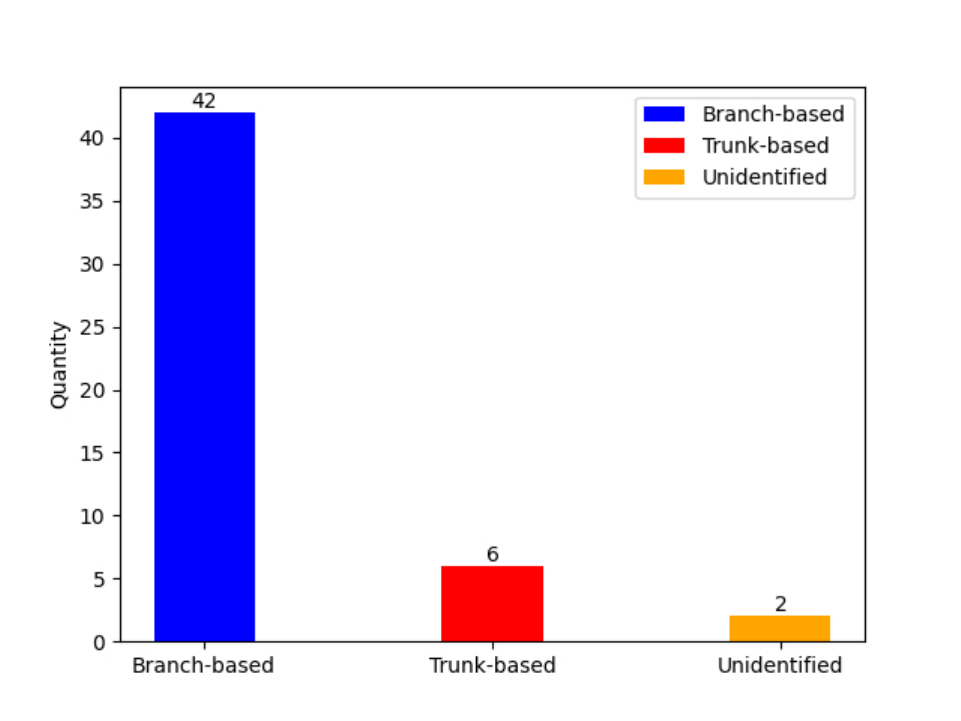}
  \Description{A bar graph showing who in our sample uses branch-based and trunk-based workflows}
\end{figure}

In Figure \ref{fig:branch_vs_trunk}, branch-based models, including GitFlow and GitHub Flow, cited three and two times, respectively, are significantly more popular than trunk-based models in our sample. Out of the total 42 responses (84\%) for branch-based, only 6 responses (12\%) were for trunk-based. Furthermore, participants cited the Stacked Diffs workflow twice, a model that involves applying a series of changes on top of each other.

\begin{figure}[hbt!]
  \centering
  \caption{\label{fig:branch_trunk_interviews} Interviews Results: Branch-based vs. Trunk-based}
  \includegraphics[width=\linewidth]{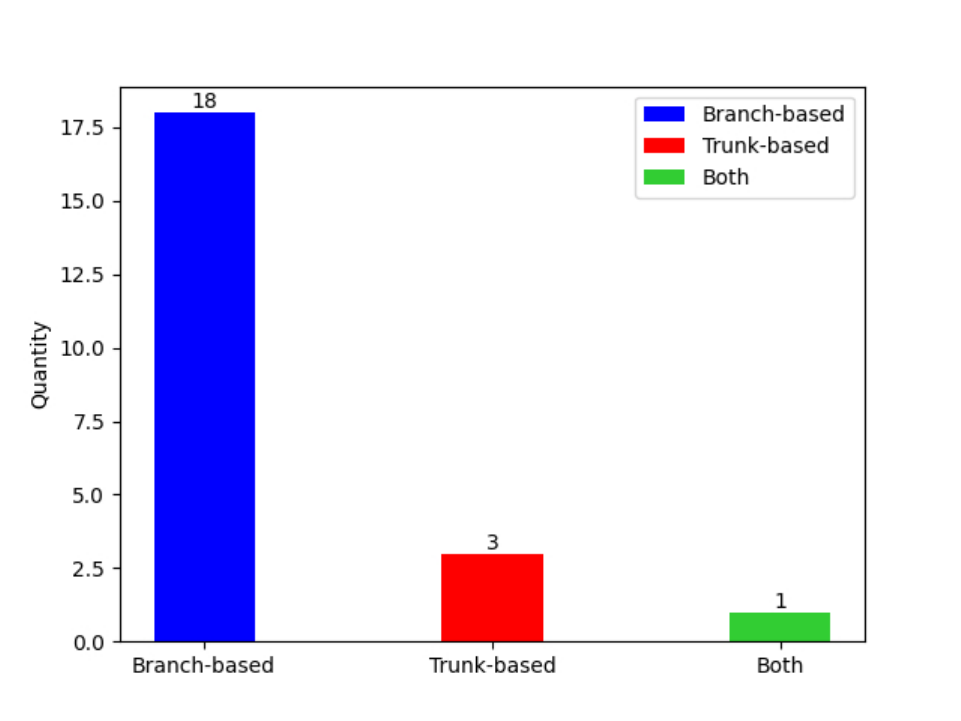}
  \Description{A bar graph showing who in our sample uses branch-based and trunk-based workflows}
\end{figure}

In the interviews, Figure \ref{fig:branch_trunk_interviews} shows the same trend as in the survey data. That is, branch-based is much more popular than trunk-based. Additionally, one participant claimed to use both workflows simultaneously, choosing between them based on task complexity: for simpler tasks such as maintenance or updates, they preferred trunk-based development, whereas for more complex or ongoing system development, they followed the GitFlow model.

The survey data was then used to conduct two more investigations, one of which was to see the impact of different demographics, such as experience and team size, on the decision-making process of employing a workflow, plotting box plots comparing how the graphs vary if we selected only those who answered a specific model.

\begin{figure}[hbt!]
  \centering
  \caption{\label{fig:experience_vs_workflow} Survey Results: Developers' Years of Experience across Workflows}
  \includegraphics[width=\linewidth]{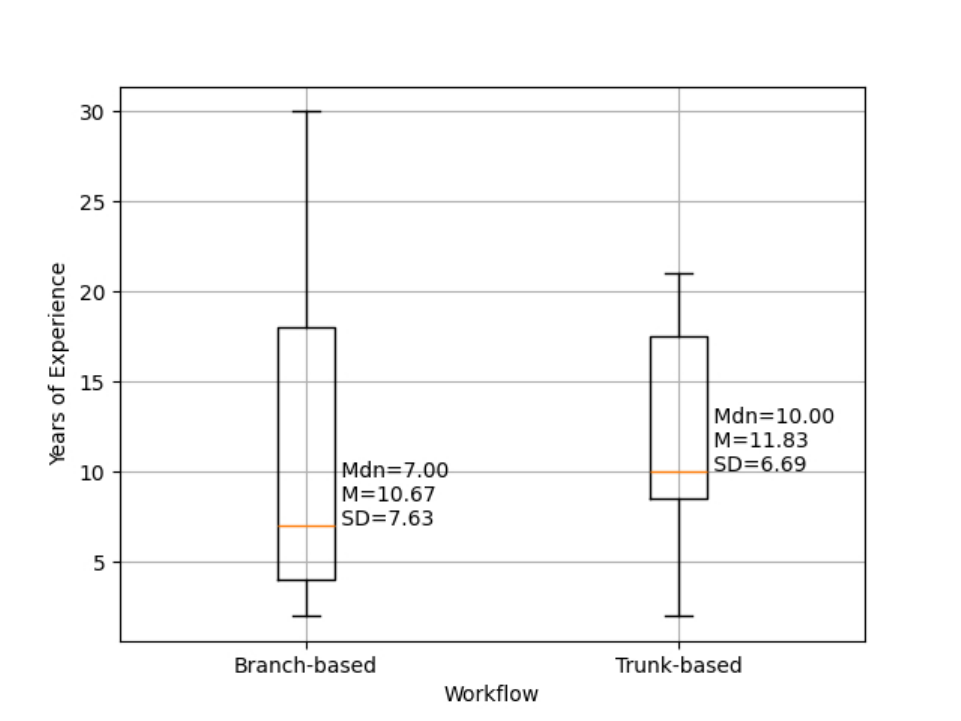}
  \Description{Two boxplots in only one graph, one for those who utilize branch-based models and one for those who utilize trunk-based models}
\end{figure}

\begin{figure}[hbt!]
  \centering \caption{\label{fig:team_size_x_workflow} Survey Results: Team Size across Workflows} \includegraphics[width=\linewidth]{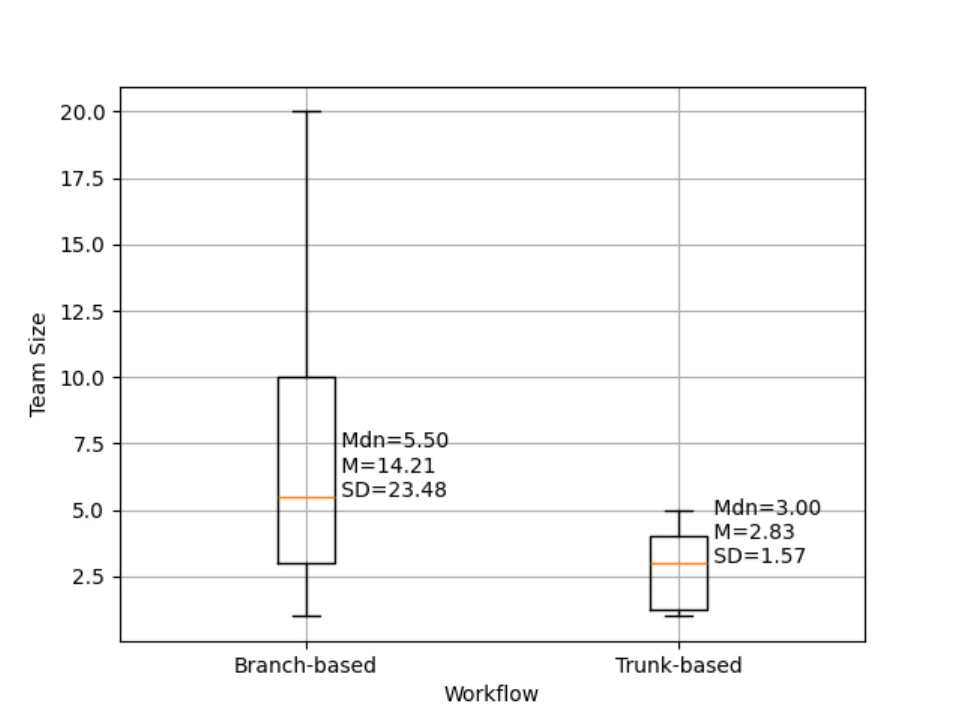}
  \Description{Two boxplots in only one graph, one for those who utilize branch-based models and one for those who utilize trunk-based models}
\end{figure}

Figure \ref{fig:experience_vs_workflow} shows the box plot that compares the experience of developers for each workflow. In the trunk-based boxplot, all answers range between 2 and 22, while the branch-based range from 2 to 30; this more considerable variation explains why the branch-based $SD$ is greater than the trunk-based $SD$. On the other hand, the trunk-based model has a higher $Mdn$ and $M$ than the branch-based model (10 and 11.83 compared to 7 and 10.67), indicating that the developers who answered that use trunk-based development are slightly more experienced than those who use the other model.

To address the second investigation, parallel to how we approach years of experience per workflow box plots in Figure \ref{fig:experience_vs_workflow}, we analyzed how the team size varies if we selected only those who responded who employ a specific model.

Figure \ref{fig:team_size_x_workflow} shows that the boxplot for branch-based models has metrics close to the ones presented in Subsection \ref{sub:yoe}, this is due to branch-based users being a large portion of our answers, as stated in Figure \ref{fig:branch_vs_trunk}.

Starting our analysis on trunk-based models, we can see in Figure \ref{fig:team_size_x_workflow} that it ranges from a single person to at most 5. In contrast, the branch-based box plot ranges from one developer to a maximum of 20. These data show a trend in our sample for trunk-based being preferred by smaller teams.

Note that in branch-based, the $SD$ is significantly higher than $Mdn$ and $M$, due to the outliers shown in Figure \ref{fig:team_size}, whereas in trunk-based, the $SD$ has a low value \rev{because, in our sample,} all teams with more than 20 developers reported employing a branch-based model. 

To ensure that the high variance observed in Figure \ref{fig:team_size_x_workflow} was not driven solely by a few cases, we performed an additional analysis that removed statistical outliers based on the Interquartile Range (IQR) method. The resulting distribution remained consistent with the original plot, reinforcing our previous findings. 

In particular, the trunk-based model continued to exhibit low dispersion and a clustered team size distribution, while the branch-based workflow retained a broader range of team sizes. In our sample, even when controlling for outliers, trunk-based workflows are typically adopted by smaller, more homogeneous teams, whereas branch-based models are employed by teams with more varied sizes.

\rev{Additionally, we conducted a preliminary analysis of statistical tests to evaluate these findings. Due to the small trunk-based group (n = 6) and imbalanced sample sizes, distributional assumptions could not be tested. To address this, we used the Mann-Whitney U test, a non-parametric method.}

\rev{Results showed a significant difference in team size, but not in years of programming experience. A power analysis indicated that the small trunk-based group considerably limits the strength of these statistical test findings (power = 32.9\%). Given the limited number of observations in the trunk-based model, we avoid drawing conclusions based on these results, as statistical tests for group differences are unreliable with such small samples.}

Furthermore, in the survey, 76.5\% of participants indicated they would not change their current project model if they could, while 3.9\% would change only if specific factors, such as an increase in the development team, occurred. Additionally, 5.9\% expressed a desire for change, and 13.7\% did not respond.

We did not conduct this comparative analysis on the data from our interviews because only three interviewees used trunk-based. However, we also asked them if they would like to change their current workflow and why. Seven participants answered yes, but they would only change some steps in their process, like adding CI/CD routines or customizing branches, instead of switching between branch-based and trunk-based workflows.

Taking into account our results, we answer \textbf{RQ1} as follows:

Regarding our sample, branch-based models seem far more popular and disseminated than trunk-based models. However, despite GitFlow being the most popular model, participants \rev{in both the survey and interviews} rarely mention it explicitly. Thus, models customized by each team are more prevalent in practice. 

Because all of our respondents are Brazilians, we cannot generalize our results so much. However, it is important to note that we do not know if our participants worked only for Brazilian companies or abroad.

Moreover, our data suggest that teams composed of more experienced developers and smaller development teams may favor trunk-based development, but further investigation is needed to assert this relationship since our sample is imbalanced. 

Finally, most participants would not like to change from trunk-based to branch-based and vice versa, although some would like to change parts of their workflow.

\subsection{RQ2. What factors favor or hinder using a branch-based or a trunk-based workflow?}
\label{sub:rq2}
In this Section, we describe the factors that our participants pointed out in both the survey and the interviews as relevant to the choice of a workflow type, \rev{organized in descending order of frequency, with each group title including the number of times the factor was cited by respondents}. Specifically, we asked which factors favor or disfavor using a specific workflow.

\subsubsection{Survey results}
\hfill\

To address RQ2 in the survey, we used online tools like Figma and the card sorting method \cite{cardsorting} to analyze its data. We extracted the central theme from our responses and created a sticker for each on the cardboard. 

In cases where related themes appeared, for example, terms like "ease" and "lower complexity" conveying the same idea, we created a group with both stickers, repeating the process until every topic was addressed and in a group. Finally, we named each group with the category that best represented the stickers.

Please note that not every sticker is part of a group. This may be because some stickers represent multiple themes at once, or there simply were not enough stickers to form a group, \rev{where at least two stickers representing the same factor were required for grouping}. However, the absence of a group does not imply that these topics are any less important.

We will cover the factors that developers answered that encourage or deter the employment of each model, with direct citations from the survey participants (SP) that explain their reasons. The card sorting sessions are available in our Git repository \cite{GIT_WORKFLOWS_REPO}.

Regarding branch-based development, the most common themes that promote this model usage were named as follows: Code Organization and Self-Containment, but other groups like Task Division, Maintenance, Conflict and Error Detection, and Flexibility, with other minor groups being present too.

\begin{enumerate}
    \item \textbf{Code Organization \rev{[6 mentions]}}: Respondents highlighted that the model has an organized commit flow, such as greater control over commits, promotes organized code practices, and reflects an organized repository structure.

    \textit{"One of the main factors is the organization of the code." (SP09)}

    \item \textbf{Self-Containment \rev{[4 mentions]}}: In this group, the answers emphasized that this model ensures parallel development, has flexible feature development, and praised the self-containment of each feature on its branch.

    \textit{"I like how I can share my tasks in a modular way by features." (SP19)}    
    \item \textbf{Task Division \rev{[3 mentions]}}: This theme stressed that this model is appropriate when being worked on by different teams.
    
    \textit{"The separation between Development and Quality Assurance teams is important." (SP10)}
    
    \item \textbf{Maintenance \rev{[3 mentions]}}: The respondents in this group emphasized that this model has facilitated rollbacks, release code maintenance, faster feature branch maintenance, and faster deployment to production

    \textit{"It makes maintenance of code that has already been released viable without altering the flow of the current and future releases." (SP20)}
    
    \item \textbf{Conflict and Error Detection \rev{[2 mentions]}}: The participants stated that the branch-based model helps avoid code conflicts and has quick error detection.

    \textit{"This model is good at avoiding merge conflicts." (SP12)}

    \textit{"This model helps merge control and avoid conflicts\ldots" (SP54)}
    
    \item \textbf{Flexibility \rev{[2 mentions]}}: This group stressed the versatility of this model and the freedom that developers have in their branches.

    \textit{"Developers can collaborate in whatever way they prefer in their branches." (SP18)}    
    
\end{enumerate}

In addition to these topics, some stickers ended up without a group because they covered many themes at the same time, like one that stated that for applications, such as mobile apps, that undergo store releases, Quality Assurance (QA) reviews, and specific fixes for release, it is more advantageous to have a branch-based model to separate what is still in development, what is in each release, and what and from which release will be fixed.

Moreover, the following groups discourage branch-based workflow usage: Complexity, Small Teams, Branching Issues and Inexperienced Teams.

\begin{enumerate}
    \item \textbf{Complexity \rev{[4 mentions]}}:  In this group, the responses emphasized that this model has a higher release complexity, a higher code alteration complexity, more documentation, and steps to release features, leading to a longer development cycle.

    \textit{"Using this model is more labor intensive." (SP20)}  
    
    \item \textbf{Small Teams \rev{[4 mentions]}}: Statements like "With fewer team members, the development will take longer due to having more steps and stages to follow".
    
    \item \textbf{Branching Issues \rev{[4 mentions]}}: Respondents highlighted that branch-based models have extensive product segmentation, that having too many branches with long lifespans can make them outdated and prone to conflicts, and that often a considerable amount of time is spent analyzing each branch.

    \textit{"If branches have a very long lifespan, they become outdated and cause conflicts." (SP18)} 

    \item \textbf{Inexperienced teams \rev{[2 mentions]}}: In this group, two statements were cited that express that branch-based models need focused and experienced teams.

    \textit{"It requires the focus and maturity of the team to maintain organization." (SP06)} 
\end{enumerate}

One sticker left without a group states that a factor that hinders this model from being employed is that maintenance of epic features can take a long time, often requiring rebases or merges with updates.

In trunk-based development, the main groups that emerged from the survey answers that promote this model were named as follows: Easiness and Practicality, Agility, Quick Versioning, Team and Project Size, and Back end and Server, besides other minor ones explained below.

\begin{enumerate}
    \item \textbf{Easiness and Practicality \rev{[5 mentions]}}: This group underlined the minor complexity, simplicity, and ease of use of the model.

    \textit{"It is easier for the team to align on the workflows to follow when there are fewer branches and steps in a remote work context." (SP08)}
    
    \item \textbf{Agility \rev{[5 mentions]}}:  The respondents in this group emphasized the model's speed, agility, and faster development process, stating that the trunk-based workflow works best with applications that need faster feedback cycles.

    \textit{"Good for applications that are small or require quick feedback." (SP10)}
    
    \item \textbf{Quick Versioning \rev{[5 mentions]}}: Themes like having a less strict workflow were covered in this group, mentioning that, generally, trunk-based models have lower complexity, with more minor changes, such as text or layout not having to pass through many tests.

    \textit{"It is easier for the team to align on the workflows to follow when there are fewer branches and steps in a remote work context." (SP08)}

    \item \textbf{Team and Project Size \rev{[5 mentions]}}: Most of the answers in this group were related to this model being good in small projects or when it has a small team.

    \textit{"This model is good when only a few people are maintaining the project." (SP01)}
    
    \item \textbf{Management \rev{[2 mentions]}}: In this case, the answers highlighted that this model has easy maintenance and a good application monitoring structure.

    \textit{"There is a good application monitoring structure." (SP10)}
    
    \item \textbf{Failures \rev{[2 mentions]}}: This group responded that trunk-based models have fewer risk factors and rarer merge failures, although the respondents did not specify such risk factors.

    \textit{"There are fewer merge failures." (SP06)}
    
    \item \textbf{Backend and Server \rev{[2 mentions]}}: The answers in this group stated that backend code works better with this model due to the less strict versioning flow and that they typically have few or only one main branch when working on server code.

    \textit{"Backend code functions better with this approach due to the less strict versioning flow." (SP08)}
\end{enumerate}

There were only two stickers without a group, one stating that this model promotes CI/CD practices, and the other stating that it favors focused and experienced teams.

However, the following groups were formed that discourage trunk-based workflows: Team and Project Size, Maintenance, Management, and some minor groups like CI/CD and Security.
 
\begin{enumerate}
    \item \textbf{Team and Project Size \rev{[7 mentions]}}: This group stated that trunk-based development could have inferior performance when the project has a lot of contributors, features, tasks, and products, and with that, the team must focus on achieving success.

    \textit{"If there are too many people committing changes simultaneously, it becomes difficult to keep the code updated." (SP08)}
    
    \item \textbf{Maintenance \rev{[5 mentions]}}: Respondents in this group cited that the model has low flexibility, interference between commits and pushes frequently, and there is no easy way to do feature toggles, meaning that in case they have to remove some feature, this can have an impact on others that are not related.

    \textit{"It is easy to encounter interference between commits/pushes." (SP18)}
    
    \textit{"Absence of application monitoring and feature toggles." (SP10)}
    
    \item \textbf{Management \rev{[2 mentions]}}: The answers highlighted that this model does not deal with application monitoring and quality assurance well.

    \textit{"Quality control becomes challenging." (SP18)}
    
    \item \textbf{CI/CD \rev{[2 mentions]}}: The themes of this group stated that respondents do not recommend this workflow with slow CI/CD pipelines and that teams with no experience in this practice may encounter difficulties.

    \textit{"This model execution is difficult in teams with little experience in CD." (SP10)}
    
    \item \textbf{Security \rev{[2 mentions]}}: In this one, developers stated that they do not like this model's low security of production code.

    \textit{"This model lacks testing security." (SP19)}    
\end{enumerate}

In addition, two stickers were not assigned because the person who responded indicated a lack of experience with trunk-based development. However, he thinks this model could often result in frequent conflicts and challenges in executing rollbacks within a production environment.

\subsubsection{Interviews results}
\hfill\

In this Section, we briefly describe ten factors that interview participants (IP) state as relevant when choosing a branch-based or a trunk-based workflow, organized from most to least cited. As mentioned previously, we used open coding to analyze the transcription of our interviews.

\begin{enumerate}
    \item \textbf{Separate Environments \rev{[17 mentions]}}: This factor considers the separation of environments, such as development, quality assurance, pre-production, and others, before the production environment, each potentially having a branch or being based on a specific commit. For example, one of the interviewees mentioned: "...reverting to a more traditional model of separated environments." (IP1), in this case, they worked in a trunk-based development structure and transitioned to branch-based, specifically GitFlow.

    \item \textbf{Code Management Complexity \rev{[12 mentions]}}: This factor considers the complexity of managing version control and how costly and complicated it can be when using branch-based models. In general, the more branches there are, the more complex it becomes to manage and ensure they are updated as they should be. For example, one of the interviewees mentioned: "...the complexity is lower (when using trunk-based workflow)..." (IP2).

    \item \textbf{Testing and Rapid Deployment to Production \rev{[11 mentions]}}: This factor considers how quickly the development team can code and effectively deploy code changes unrelated to incidents or bug fixes to production. We have an example from one of the interviewees who said: "...the advantage is because we have fewer processes when deploying to production because it is just one pipeline..." (IP10); in this case, he was working in a trunk-based Development structure, \rev{transitioned to GitFlow, and after the migration, noticed the trunk-based model's rapid deployment to production as a factor}. Therefore, branch-based workflows suffer in this regard due to the branch management processes, often resulting in costly and slow deployments.

    \item \textbf{Rigid Processes \rev{[10 mentions]}}: This factor takes into account management control. Whether there is any degree of freedom to decide which workflow to use, or if a process is mandatory and standard. For example, we have the quote of one of the interviewees who said: "... For us, where we work, the worst thing is dealing with bureaucracy ..." (IP18). Therefore, in environments where the process is more important than the delivery itself, branch-based workflows tend to be more popular at the expense of trunk-based ones.

    \item \textbf{Teams composed mostly of less experienced developers \rev{[8 mentions]}}: This factor considers teams in which the interviewee explicitly mentioned the aspect of lower seniority among team members. For example, one of the interviewees said: "...because in all projects, we have multidisciplinary teams with various levels of knowledge, right? So basically, in current projects, we have many developers who are not senior..." (IP15). Participants claimed that less experienced developers might be hindered from using a trunk-based workflow, whereas branch-based workflows provide them with greater security.

    \item \textbf{Focus on Quality \rev{[7 mentions]}}: This factor considers the explicit concern for code quality, even though it should be implicit in any implementation, workflow, and so on, as quality is one of the primary requirements for any development. For example, one of the interviewees said: "... It is a cool thing (the branch-based workflow). Even for code improvement reasons..." (IP9). Therefore, participants frequently related branch-based workflows to code quality.

    \item \textbf{Autonomy \rev{[7 mentions]}}: This factor considers the freedom to work independently from what others in the team are doing, being one of the basic precepts of Git and even expected to appear favoring branch-based workflows. For example, one of the interviewees said: "And then you get a certain independence to work..." (IP3).

    \item \textbf{Fast Incident Response Time \rev{[4 mentions]}}: This factor considers how quickly it is to deploy a fix to a production incident. For example, one of the interviewees said: "It (the trunk-based workflow) was much faster; in less than half a day, the fix was already in production." (IP1). Participants reported this factor as disfavoring the use of branch-based workflows.

    \item \textbf{Focus on Delivery Speed \rev{[4 mentions]}}: This factor considers fast deliveries, where smaller portions of code are delivered, favoring trunk-based workflows. For example, one of the interviewees said, "Our project is quite fast, so there are features that are small, and they go up (into production) quite frequently..." (IP13).

    \item \textbf{Restricted Sector \rev{[4 mentions]}}: This factor considers the type of sector in which the interviewee works, such as the financial sector, which has its particularities, like a greater focus on security, favoring branch-based workflows. For example, one interviewee mentioned: "...a company in the financial sector..." (IP5).
\end{enumerate}

\subsubsection{RQ2 results discussion}
\hfill\

The survey results align with the interview findings, reinforcing the evidence gathered in this study. Furthermore, the trend shown in \textbf{RQ1} that smaller teams and developers with higher levels of experience are characteristics that favor the use of trunk-based workflows is complemented and reinforced by the responses in \textbf{RQ2}.

While branch-based development can be flexible and offer self-containment in the form of feature branches, leave the code organized, and promote task division and code quality, there are still challenges, whether due to its complexity or because it overly segments the final product into numerous branches, making the integration and deployment process to be often costly, and delaying releases and feedback cycles.

On the other hand, trunk-based development promotes agile development and CI/CD practices with simple and quick versioning. However, it requires a focused development environment, which can be challenging for larger teams or teams that need to work with the same repository, and those without experienced developers due to the maintenance difficulty.

It is interesting to note developers' perceptions of conflicts occurring in both workflows. Some participants indicated that branch-based workflows are prone to conflicts because long-lived branches may diverge more from the main branch, causing more severe conflicts. In contrast, some participants pointed out that the trunk-based workflow is prone to conflicts, especially when team developers have a lower level of seniority.

Nevertheless, both workflows have their merits and are entirely consistent with the autonomy factor, as both provide freedom for the developer to work independently from others. The choice between them depends on the project needs, team culture, and practical development practices to manage associated challenges.

To choose which type of workflow to use, the developer can use all the themes and factors listed in this study as a checklist and choose which has the most characteristics in common with their specific context.

For example, a startup context, where the product is simple, the development team is lean, and there is a need for quick deliveries and customer feedback, would benefit from a trunk-based workflow.

In contrast, the context of a large company, with multiple teams working on the same repository, where there is high developer turnover with different levels of experience, would benefit from the self-containment of a branch-based workflow in which each branch represents a separate environment.
\section{Threats to Validity}\label{sec:threats}
\label{cap:threats}

In this Section, we will discuss the possible factors that might represent threats to this work's validity, \rev{which we organize following the classification by Wohlin et al. \cite{WOHLIN}}.

\rev{\textit{Construct Validity}. This study may be threatened by subjectivity in classifying workflows from open-ended responses, as well as by differences between survey and interview instruments}. Another potential threat is our own bias towards one workflow or another, stemming from previous experiences. To mitigate this issue, we carefully designed our survey and interview questions to remain neutral and avoid references to specific factors that could influence participants’ responses. Furthermore, our analysis relied exclusively on participants' statements, avoiding external interpretations.

\rev{\textit{Internal Validity}. Potential confounding factors, such as differences in development methodologies or company policies, were not controlled}. \rev{Additionally, there is a risk of receiving multiple responses from individuals within the same company or team, which could skew results. To mitigate this, the survey was distributed via general mailing lists and social media, rather than targeting specific companies or environments where employees might be encouraged to participate}.

\rev{\textit{External Validity}}. Our sample is composed entirely of Brazilian developers, even though some of them work for foreign companies, which may limit the generalizability of our findings to other contexts. Furthermore, even though we used general mailing lists, participation was voluntary, and our invitation was unsolicited, which might have disturbed their daily activities. Unfortunately, there are no dedicated portals or guidelines for gathering volunteers for software engineering research. To mitigate this, we designed a short and cordial invitation email that also acted as an informed consent form, including ethical and privacy considerations, and made participation entirely opt-in \cite{SPAM, challenges_survey}.

\rev{\textit{Conclusion Validity}. Our sample imbalance, especially the low number of trunk-based workflow users, limits the statistical strength of comparisons}.
\section{Related Works}\label{sec:relatedworks}

The challenge of maximizing team productivity while promoting software quality has led to significant research on collaborative software development workflows.

Studies delved into developers' practices and their effects on productivity and code quality. Shihab et al. \cite{SHIHAB} analyzed data from two releases of major Microsoft projects, finding that team organization directly influences quality and that branching affects post-release failures. They suggest that aligning branch structure with team organization slightly reduces adverse effects on software quality compared to architecture-based structures. While their research focuses on a single corporate environment, our work amplifies this approach by including various companies, methodologies, and workflows.

Bird and Zimmermann \cite{BIRD} conducted a survey on branch usage in a large Microsoft project, identifying common issues. They performed a what-if analysis to evaluate alternative branch structures based on isolation and liveness. By removing low-benefit, high-cost branches, they found potential savings of 8.9 days in delays with only 0.04 additional conflicts on average. A significant issue highlighted was the lengthy delays in changes moving between teams, often due to excessive branches. In our work, we identified extensive product segmentation as a factor hindering the usage of branch-based models.

Costa et al. \cite{COSTA} surveyed 109 software developers to understand their use of branches, integration challenges, and strategies to minimize issues. They discovered that developers mainly use branches for new features or bug fixes. According to the participants, developers can freely create new branches, but conflicts often arise due to prolonged isolation of branches and poor communication. In our work, some interviewees mentioned long-lived branches as a factor that discourages using a branch-based workflow.

Phillips et al. \cite{branching_and_merging} recruited 140 version control users to participate in an online survey to collect information on branching and merging practices in software development projects. The findings claim that in their sample, the types of branches created influence developers' branching satisfaction, with feature, release, and experimental branches having the most impact. Furthermore, staged (branch-based) topologies are increasingly more common in their sample when filtering the data by project size. Their data also shows that branch-per-release or features were the most popular among respondents. In our work, we also observed this pattern.

\rev{Studies compared workflows in a specific development context}. De Boer \cite{DEBOER} conducted a study in a large software company transitioning from trunk-based to branch-based development with merge requests. Through interviews and surveys with developers before and after the migration, the study identified that trunk-based development favored delivery speed. In contrast, the branch-based model improved perceived code quality, despite some integration challenges. Our study expands on this perspective by comparing multiple companies with different workflows and identifying additional influencing factors, such as Separate Environments, which the literature has not previously addressed.

\rev{Neely \& Stolt \cite{NEELYSTOLT} present an experience report on a company transition from time-boxed Scrum sprints to continuous delivery with Kanban, highlighting technical and organizational challenges. Their experience emphasizes factors that align with our findings about trunk-based workflows, such as agility, faster feedback cycles, and reduced complexity in versioning. Similarly, our works indicate the need for test automation and monitoring structures to mitigate risks. Challenges related to maintenance, coordination among teams, and the importance of organizational commitment also appear in our studies, but similarly to \cite{DEBOER}, their work focused on a single organizational context}.

Cortés Ríos et al. \cite{CORTESRIOS} extracted characteristics such as branching strategies from workflows in public Git repositories through literature analysis and README file mining. Their framework categorized these workflows, revealing that claimed similarities often masked significant differences. However, they did not evaluate the practical advantages and disadvantages of the various workflows. Future research could explore whether our respondents' workflow descriptions align with this framework.

Rayana et al. \cite{RAYANA} introduced GitWaterFlow, a flexible branching model that supports long-term maintenance of older product versions while enabling quick deployment of critical bug fixes. Unlike GitFlow, it uses atomic multi-branch merging, allowing developers to integrate multiple fixes or updates. Their team's experience using the GitWaterFlow model for nearly a year suggests that the model has led to faster deliveries and a general reduction in engineering work due to the new deployment method, successfully meeting their goals in the given setting, with room for fine-tuning and improvement of processes and tools.

\hfill
\section{Conclusions}\label{sec:conclusions}
This work aimed to understand how developers work with Git, considering branch-based and trunk-based workflows, and which factors favor or hinder using a specific workflow. To this end, we conducted survey research with 54 developers and semi-structured interviews with 22 individuals.

Our results indicate that developers still use branch-based workflows more frequently than trunk-based ones. Moreover, there is a preference for trunk-based models among more experienced programmers and small development teams. However, further investigation is required to confirm this relationship due to the significant imbalance in our sample (only 12\% employing trunk-based models).

Also, the factors developers cited that promote or hinder the usage of branch-based workflows is that while it can be flexible, offer self-containment in the form of feature branches, leave the code organized, and promote task division, there are still challenges, whether due to its complexity or because it overly segments the final product into numerous branches.

On the other hand, trunk-based development promotes agile development. Allowing CI/CD practices and having simple and quick versioning, but demanding a focused development environment in exchange, can be challenging for teams without experienced developers because of the maintenance difficulty.

Finally, both workflows have their advantages, and choosing the most suitable one depends on various circumstances, such as the project size, necessities, the seniority of the developer team, and the development guidelines adopted by the company.
\section*{Artifacts Availability}
All data collected through our survey and interviews are available in this repository \cite{GIT_WORKFLOWS_REPO}.
\begin{acks}
\rev{We thank the participants of our survey and interviews. For partially supporting this work, we would like to thank INES (National Software Engineering Institute) and the Brazilian research funding agencies CNPq, FACEPE, and CAPES}.
\end{acks}

\bibliographystyle{ACM-Reference-Format}
\bibliography{sample-base}

\end{document}